\documentclass[12pt,a4paper]{article}
\usepackage{a4wide,cite,axodraw}

\newcommand{\1}[1]{\mathbf{1}^{\mathbf#1}}
\newcommand{\2}[1]{\mathbf{2}^{\mathbf#1}}
\newcommand{\3}[1]{\mathbf{3}^{\mathbf#1}}
\newcommand{\4}[1]{\mathbf{4}^{\mathbf#1}}
\newcommand{\0}[1]{\mathbf{0}^{\mathbf#1}}
\newcommand{\Li}{\mathop{\mathrm{Li}}\nolimits_2}
\newcommand{\tfrac}[2]{{\textstyle\frac{#1}{#2}}}
\newcommand{\ep}{\varepsilon}
\begin{document}

\begin{titlepage}

\begin{flushright}
{MZ-TH/98-38}\\[3mm]
{hep-ph/9809589}\\[5mm]
{September 1998}\\
\end{flushright}

\vspace{3cm}

\begin{center}
{\Large\textbf{Effect of $m_c$ on $b$ quark chromomagnetic
interaction\\[2mm]
and on-shell two-loop integrals with two masses}}
\end{center}

\vspace{1cm}

\renewcommand{\thefootnote}{\fnsymbol{footnote}}

\begin{center}
A.~I.~Davydychev\footnote{Alexander von Humboldt fellow.
On leave from the Institute for Nuclear Physics, Moscow State
University, 119899 GSP Moscow, Russia.
Email: davyd@thep.physik.uni-mainz.de}
\ and
A.~G.~Grozin\footnote{Permanent address:
Budker Institute of Nuclear Physics,
Novosibirsk 630090, Russia.
Email: A.G.Grozin@inp.nsk.su}\\[1cm]
\textit{Department of Physics, University of Mainz,\\ 
Staudinger Weg 7, D-55099 Mainz, Germany}
\end{center}

\renewcommand{\thefootnote}{\arabic{footnote}}
\setcounter{footnote}{0}

\vspace{1cm}

\begin{abstract}
The effect of non-zero $c$ quark mass on $b$ quark HQET Lagrangian,
up to $1/m_b$ level, is calculated at two loops.
The results are expressed in terms of dilogarithmic functions
of $m_c/m_b$.
This calculation involves on-shell two-loop propagator-type diagrams 
with two different masses, $m_b$ and $m_c$.
A general algorithm for reducing such Feynman integrals
to the basis of two nontrivial and two trivial integrals
is constructed.
\end{abstract}

\end{titlepage}

\setcounter{page}{2}

\section{Introduction}
\label{Sec:Intro}

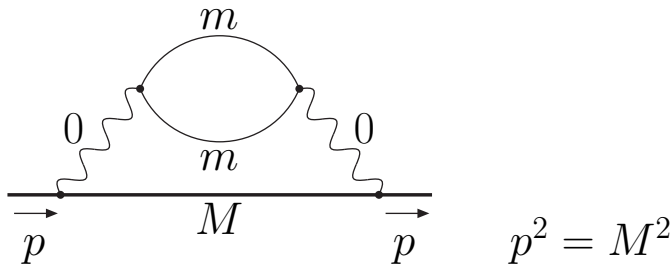
\begin{figure}[ht]
\begin{center}
\begin{picture}(160,90)
\SetWidth{1.3}
\Line(0,20)(160,20)
\SetWidth{0.5}
\Photon(20,20)(50,60){4}{3.5}
\Photon(110,60)(140,20){4}{3.5}
\CArc(80,47.5)(32.5,22.62,157.38)
\CArc(80,72.5)(32.5,202.62,337.38)
\Vertex(20,20){1.5}
\Vertex(140,20){1.5}
\Vertex(50,60){1.5}
\Vertex(110,60){1.5}
\LongArrow(2.5,13)(17.5,13)
\LongArrow(142.5,13)(157.5,13)
\Text(10,-6)[b]{\Large$p$}
\Text(80,17)[t]{\Large$M$}
\Text(150,-6)[b]{\Large$p$}
\Text(80,37)[t]{\Large$m$}
\Text(80,83)[b]{\Large$m$}
\Text(25,45)[]{\Large$0$}
\Text(135,45)[]{\Large$0$}
\Text(190,-6)[bl]{\Large$p^2=M^2$}
\end{picture}
\end{center}
\caption{Two-loop self-energy diagram.}
\label{Fig:Self}
\end{figure}

On-shell two-loop calculations in any theory
containing two massive fields with different masses
and a massless field
necessarily involve diagrams like Fig.~\ref{Fig:Self}.
Such diagrams appear, for example, 
in QED with $e$, $\mu$ and $\tau$,
in QCD with $b$ and $c$ quarks,
and in the electroweak theory.
In general, such diagrams have a two-particle threshold\footnote{It 
coincides with the corresponding pseudothreshold, since one of the
particles involved is massless.}
at $p^2=M^2$, and a three-particle threshold at $p^2=(M+2m)^2$,
where $p$ is the external momentum.
There are two three-particle pseudothresholds, at $p^2=(M-2m)^2$
and $p^2=M^2$. Therefore, the on-shell condition $p^2=M^2$
means that we are, at the same time, at the two-particle 
(pseudo)threshold and at the three-particle pseudothreshold.

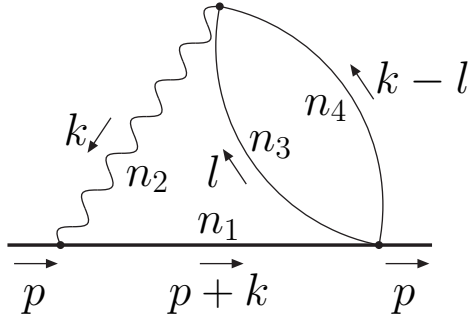
\begin{figure}[ht]
\begin{center}
\begin{picture}(160,110)
\SetWidth{1.3}
\Line(0,20)(160,20)
\SetWidth{0.5}
\Photon(20,20)(80,110){4}{6.5}
\CArc(65,35)(76.4853,-11.31,78.69)
\CArc(155,95)(76.4853,168.69,258.69)
\Vertex(20,20){1.5}
\Vertex(140,20){1.5}
\Vertex(80,110){1.5}
\LongArrow(72.5,13)(87.5,13)
\LongArrow(2.5,13)(17.5,13)
\LongArrow(142.5,13)(157.5,13)
\LongArrow(39,68)(31,56)
\LongArrow(90,43)(82,55)
\LongArrow(138,75)(130,87)
\Text(80,-6)[b]{\Large$p+k$}
\Text(10,-6)[b]{\Large$p$}
\Text(150,-6)[b]{\Large$p$}
\Text(26,64)[]{\Large$k$}
\Text(78,48)[]{\Large$l$}
\Text(157,83)[]{\Large$k-l$}
\Text(80,22)[b]{\Large$n_1$}
\Text(53,45)[]{\Large$n_2$}
\Text(99.5,58)[]{\Large$n_3$}
\Text(120.5,72)[]{\Large$n_4$}
\end{picture}
\end{center}
\caption{On-shell two-loop integral with two masses.}
\label{Fig:Int}
\end{figure}

Combining the identical massless denominators,
we can express these diagrams via scalar integrals
(see Fig.~\ref{Fig:Int})
\begin{equation}
I(n_1,n_2,n_3,n_4) \equiv - \frac{1}{\pi^d} \left.
\int \frac{\mathrm{d}^d k\, \mathrm{d}^d l}
{\mathcal{D}_1^{n_1}\mathcal{D}_2^{n_2}
\mathcal{D}_3^{n_3}\mathcal{D}_4^{n_4}}\right|_{p^2=M^2} \,,
\label{I}
\end{equation}
where
\begin{equation}
\label{D_i}
\mathcal{D}_1 = M^2 - (p+k)^2\,,\quad
\mathcal{D}_2 = -k^2\,,\quad
\mathcal{D}_3 = m^2-l^2\,,\quad
\mathcal{D}_4 = m^2-(k-l)^2\,,
\end{equation}
and $d=4-2\ep$ is the space-time dimension
(in the framework of dimensional regularization~\cite{dimreg}).
Evidently, 
\[ 
I(n_1,n_2,n_4,n_3)=I(n_1,n_2,n_3,n_4)\; . 
\]

Some integrals of this class have been already considered in the 
literature~\cite{El,GBGS,BGS,BG,CM,FT,BDU}\footnote{The asymptotic
expansion of similar on-shell integrals has been constructed
in~\cite{CS,AK}.}.  
Here we propose a general algorithm,
based on integration by parts~\cite{CT},
which reduces any integral~(\ref{I}) (with integer $n_i$) 
to the basis of two nontrivial integrals
(known near four dimensions up to finite terms in $\ep$)
and two trivial ones.

In realistic calculations, we also get integrals~(\ref{I})
with scalar numerators.
In some cases, these numerators cannot be directly
expressed in terms of the denominators~(\ref{D_i}).
Such integrals are denoted as $\widetilde{I}(n_1,n_2,n_4,n_3;n_0)$
(where $n_0$ is the power of the corresponding numerator, see 
eqs.~(\ref{tildeI})--(\ref{N})),
and they are considered in Sec.~\ref{Sec:Int}.
They can be reduced to the same basis as the integrals
$I(n_1,n_2,n_4,n_3)$.

The integration-by-parts technique~\cite{CT} was first developed
in the context of massless calculations.
Later it was used in on-shell two-loop calculations
with a single mass~\cite{GBGS,BGS,B,FT2}
as well as in HQET~\cite{BG2} (see a short review~\cite{BG3}).

Recently it was shown~\cite{T} 
(cf.\ also \cite{GhY})
that any two-loop propagator integral
with generic masses and external momentum can be reduced
to a finite set of basis integrals\footnote{The algorithm
of~\cite{T} was originally implemented in \textsf{FORM}
and then~\cite{tarcer} in \textsf{Mathematica}.}.
However, in special cases like the thresholds and pseudothresholds
some general formulae of ref.~\cite{T} become degenerate.
Therefore, these cases require a special examination. 
In particular, 
here we propose a reduction algorithm for the integrals~(\ref{I})
and implement it as a \textsf{REDUCE}~\cite{H,G} package.

Integrals of this class appear in matching QCD
to Heavy Quark Effective Theory
(HQET, see~\cite{N} for review and references).
Specifically, let us consider the $b$ quark HQET with $m_b\equiv M$,
keeping one more massive flavour, $c$, with $m_c\equiv m$.
Coefficients in the HQET Lagrangian
(as well as in the $1/M$ expansion of QCD operators)
are derived by equating the on-shell matrix elements in QCD and HQET.
In order to determine coefficients in the Lagrangian
up to the $1/M$ level, it is sufficient to equate
scattering amplitudes of an on-shell heavy quark
in an external gluon field:
\begin{eqnarray}
&&\bar{u}(p') t^a \left( F_1(q^2) \frac{(p+p')^\mu}{2M}
+ G_{\mathrm{m}}(q^2) \frac{[\rlap/q,\gamma^\mu]}{4M} \right) u(p)
\nonumber\\
&&\quad{} = \bar{u}_v(q) t^a \left( v^\mu + \frac{q^\mu}{2M}
+ C_{\mathrm{m}}(\mu) \widetilde{Z}^{-1}(\mu)
\widetilde{\mu}_{{\mathrm{g}}0}
\frac{[\rlap/q,\gamma^\mu]}{4M}
\right) u_v(0) + \mathcal{O}\left(\frac{q^2}{M^2}\right)\,.
\label{match}
\end{eqnarray}
Here $p=Mv$, $q=p'-p$, $p^2=p^{\prime2}=M^2$;
form factors on the QCD side are $F_1(q^2)=1+\mathcal{O}(q^2)$,
$G_{\mathrm{m}}(q^2)=\mu_{\mathrm{g}}+\mathcal{O}(q^2)$;
$C_{\mathrm{m}}(\mu)$ is the coefficient
of the chromomagnetic interaction,
the only nontrivial coefficient in the HQET Lagrangian to this order;
$\widetilde{Z}^{-1}(\mu)\widetilde{\mu}_{{\mathrm{g}}0}$ 
is the matrix element
of the renormalized HQET chromomagnetic operator
(note that $C_{\mathrm{m}}(\mu)\widetilde{Z}^{-1}(\mu)$
does not depend on the normalization scale $\mu$);
the spinors are related by the Foldy-Wouthuysen transformation
$u(Mv+k)=\bigl(1+\rlap/k/(2M)+\mathcal{O}(k^2/M^2)\bigr)u_v(k)$
(see~\cite{G2} for a recent short review of the status
of calculations of coefficients in the HQET Lagrangian).
Two-loop anomalous dimension of the HQET chromomagnetic operator
(or $\widetilde{Z}$) was calculated in~\cite{ABN,CG}.
The chromomagnetic coefficient $C_{\mathrm{m}}(\mu\sim M)$
was calculated at two loops in~\cite{CG},
under the assumption that all other flavours
(except the heavy flavour of HQET) are massless;
higher orders in the large $\beta_0$ limit were summed in~\cite{GN}.

Here we calculate the effect of non-zero $c$ quark mass
on the chromomagnetic interaction in the $b$ quark HQET.
The HQET on-shell loop integrals without massive quark loops
contain no scale and hence vanish. Therefore, only the integrals
involving $c$ quark loops are relevant.
A simple method for calculating such integrals
was proposed in~\cite{BG} (and used in~\cite{CG}).
The QCD calculation involves the integrals~(\ref{I}).
The non-zero $m_c$ effect on heavy-light bilinear quark currents
in HQET was considered in~\cite{BG}.

\section{On-shell two-loop integrals with two masses}
\label{Sec:Int}

Integrals~(\ref{I}) with $n_3\le0$ or $n_4\le0$
reduce to products of one-loop ones,
and are proportional to $T_1\equiv I(1,0,1,0)$
with rational coefficients.
Integrals with $n_1\le0$ are two-loop vacuum ``bubbles'',
an explicit formula for them can be found in~\cite{GBGS};
they are proportional to $T_0\equiv I(0,0,1,1)$ with rational 
coefficients.
These trivial basis integrals are
\begin{equation}
T_0 = (m^2)^{d-2} \Gamma^2(1-d/2)\,, \quad
T_1 = (Mm)^{d-2} \Gamma^2(1-d/2)\,.
\label{T}
\end{equation}
Explicit general formulae for similar integrals with numerators
can be found in ref.~\cite{DT}\footnote{Note that
$I(-1,n_2,n_3,n_4)=I(0,n_2-1,n_3,n_4)$.}.

When $n_1$, $n_3$ and $n_4$ are all positive, we employ the
integration-by-parts technique~\cite{CT}.
Applying the operator $\frac{\partial}{\partial k}\cdot(k-l)$
to the integrand of~(\ref{I})
and substituting $p\cdot l\to (p\cdot k)(k\cdot l)/k^2$
in the numerator,
we obtain the recurrence relation
\begin{equation}
4m^2 n_4\4+ I = \Bigl[ -2d+n_1+2n_2+4n_4 - (n_1+2n_2)(\3--\4-)\2+
+ n_1(\2--\3-+\4-)\1+ \Bigr] I\,,
\label{r1}
\end{equation}
where, for example,
$\4\pm I(n_1,n_2,n_3,n_4)=I(n_1,n_2,n_3,n_4\pm1)$, etc.
Due to the symmetry, we always may assume that $n_4\ge n_3$;
the index $n_4$ can be reduced down to $n_4=1$ by~(\ref{r1}).
After that, we are left with $I(n_1,n_2,1,1)$, plus trivial
integrals which can be expressed in terms of (\ref{T}). 

Next we are going to reduce $n_1$.
To this end, we consider recurrence relations obtained by applying
$\frac{\partial}{\partial k}\cdot k$
and $\frac{\partial}{\partial k}\cdot p$
to the integrand of~(\ref{I}):
\begin{equation}
\Bigl[ d-n_1-2n_2-n_4 - n_1\2-\1+ - n_4(\2--\3-)\4+ \Bigr] I = 0\,,
\label{r2}
\end{equation}
\begin{equation}
\Bigl[ -2n_1+2n_2+n_4 + 2n_1(\2-+2M^2)\1+ - (2n_2+n_4)\1-\2+
+ n_4(\1-\3-\2+-\1-+\2--\3-)\4+ \Bigr] I = 0\,.
\label{r3}
\end{equation}
We express $\2-\4+I$ from~(\ref{r2}) and $\1-\4+I$
from its $\1-\2+$ shifted version,
and substitute both into~(\ref{r3}):
\begin{equation}
\Bigl[ d-2n_1-1 - (d-n_1-1)\1-\2+ + n_1(\2-+4M^2)\1+ \Bigr] I = 0\,.
\label{r4}
\end{equation}
Now we express $\4+I$ from~(\ref{r1}) and substitute it into
the $\2+$ shifted version of~(\ref{r2});
adding~(\ref{r4}) and substituting $n_3=n_4=1$, we arrive at
\begin{eqnarray}
4(M^2-m^2)n_1\1+ I(n_1,n_2,1,1) 
&{}={}& \Bigl[ -3d+3n_1+2n_2+5 + (d-n_1-1)\1-\2+
\nonumber\\
&&{} - 4m^2(\3-\4++d-n_1-2n_2-3)\2+ \Bigr] I(n_1,n_2,1,1) \,.
\label{r5}
\end{eqnarray}
This relation allows one to reduce $n_1$ down to $n_1=1$
(note that the term with $\3-\4+$ is trivial here).
Thus we are left with $I(1,n_2,1,1)$.

Finally, substituting $\1+I$ from~(\ref{r5})
and its shifted version $\2-\1+I$ into~(\ref{r4}),
we obtain at $n_1=n_3=n_4=1$
\begin{eqnarray}
&&\Bigl[ (3d-2n_2-6)\2- + 4(M^2+m^2)(2d-2n_2-5)
+ 16M^2m^2(d-2n_2-4)\2+ \Bigr] I(1,n_2,1,1)
\nonumber\\
&&\qquad{} = \Bigl[ (d-2)\1-(4m^2\2++1) - 4m^2\3-\4+(4M^2\2++1)
\Bigr] I(1,n_2,1,1) \,.
\label{r6}
\end{eqnarray}
This relation allows one to lower or raise $n_2$
(note that the integrals on the right-hand side of eq.~(\ref{r6})
are trivial).
Therefore, all integrals~(\ref{I}) can be expressed via 
\begin{equation}
I_0\equiv I(1,0,1,1)\, , \hspace{10mm} 
I_1\equiv I(1,1,1,1) \; ,
\label{Idef}
\end{equation}
and the trivial integrals $T_{0,1}$~(\ref{T}),
exactly in $d$ dimensions.
The basis integrals~(\ref{Idef}) in $d$ dimensions can be expressed
via hypergeometric functions ${}_3F_2$, see Appendix~A.

Expressions for $I_{0,1}$ expanded in $\ep$ up to the finite terms
can be, with some efforts, extracted from any two independent
integrals out of those calculated in~\cite{El,GBGS,BGS,BG,CM,FT}.
Some details of this procedure are discussed in the Appendix~B.
Introducing a dimensionless variable 
\begin{equation}
r \equiv m/M \; 
\end{equation}
and dilogarithmic functions\footnote{We adopt the notation $L_{\pm}$
which has been used in ~\cite{GBGS,BGS,BG} (cf., e.g., eq.~(A3)
of \cite{BG}).
In~\cite{CM}, the functions $L_{-}$ and $L_{+}$ were called
$R_1$ and $R_2$, respectively.
In~\cite{BDU} the functions
$T^+(r)=-2L_-+\tfrac{1}{2}\log^2 r+\tfrac{1}{3}\pi^2$ and
$T^-(r)=-2L_++\tfrac{1}{2}\log^2 r-\tfrac{1}{6}\pi^2$
were used. They have a nice property $T^\pm(r^{-1})=-T^\pm(r)$,
and they are analytic continuations of each other at
$r\leftrightarrow -r$.}
\begin{equation}
L_+ = - \Li(-r) + \tfrac{1}{2}\log^2 r - \log r\,\log(1+r) 
- \tfrac{1}{6}\pi^2
= \Li(-r^{-1}) + \log r^{-1}\,\log(1+r^{-1})\,,
\label{L+}
\end{equation}
\begin{eqnarray}
L_- &=& \Li(1-r) + \tfrac{1}{2}\log^2 r + \tfrac{1}{6}\pi^2
= - \Li(1-r^{-1}) + \tfrac{1}{6}\pi^2
\nonumber\\
&=& - \Li(r) + \tfrac{1}{2}\log^2 r - \log r\,\log (1-r) 
+ \tfrac{1}{3}\pi^2
\quad(r\le1)
\nonumber\\
&=& \Li(r^{-1}) + \log r^{-1}\,\log(1-r^{-1}) \quad(r\ge1)\,,
\label{L-}
\end{eqnarray}
the results for the integrals $I_{0,1}$ can be presented as
\begin{eqnarray}
\frac{I_0}{\Gamma^2(1+\ep)} &{}={}&
{}- M^{2-4\ep} \left[ \frac{1}{2\ep^2} + \frac{5}{4\ep}
+ 2(1-r^2)^2(L_++L_-) - 2\log^2 r + \frac{11}{8} \right]
\nonumber\\
&&{} - m^{2-4\ep} \left[ \frac{1}{\ep^2}
+ \frac{3}{\ep} - 2\log r + 6 \right] + \mathcal{O}(\ep)\,,
\label{I0} \\ 
\frac{I_1}{\Gamma^2(1+\ep)} &=&
M^{-4\ep} \left[ \frac{1}{2\ep^2} + \frac{5}{2\ep}
+ 2(1+r)^2 L_+ + 2(1-r)^2 L_- - 2\log^2 r + \frac{19}{2} \right]
+ \mathcal{O}(\ep)\,.
\hspace{7mm} 
\label{I1}
\end{eqnarray}
Eq.~(\ref{I0}) corresponds to a special case of more general
results for the sunset-like diagrams presented in \cite{BDU}.
Eq.~(\ref{I1}) is equivalent to the result presented in eq.~(29)
of ref.~\cite{CM}. In Appendix~B, we discuss different ways
to derive it.
Note that
\begin{equation}
L_++L_-=\tfrac{1}{2}\Li(1-r^2)+\log^2 r+\tfrac{1}{12}\pi^2
=-\tfrac{1}{2}\Li(1-r^{-2})+\tfrac{1}{12}\pi^2\,.
\label{LL}
\end{equation}

When dealing with diagrams like Fig.~\ref{Fig:Self} with numerators
and reducing them to scalar integrals,
one cannot directly express $l\cdot p$ via $\mathcal{D}_{1\ldots4}$.
Therefore, integrals similar to~(\ref{I})
but containing some extra numerators\footnote{These numerators
are not really irreducible, like e.g.\ in the three-point
functions~\cite{UD4}, since the corresponding integrals can
be reduced algebraically.} appear.
Let us define 
\begin{equation}
\widetilde{I}(n_1,n_2,n_3,n_4;n_0)
\equiv
- \frac{1}{\pi^d} \left.
\int \frac{\mathrm{d}^d k\, \mathrm{d}^d l \;\; \mathcal{N}^{n_0}}
{\mathcal{D}_1^{n_1}\mathcal{D}_2^{n_2}
\mathcal{D}_3^{n_3}\mathcal{D}_4^{n_4}}\right|_{p^2=M^2} \,.
\label{tildeI}
\end{equation}
As compared to~(\ref{I}), it contains 
an extra factor $\mathcal{N}^{n_0}$ in the numerator,
with $n_0\ge0$ and 
\begin{equation}
\label{N}
\mathcal{N} = (2l-k)\cdot p \; .
\end{equation}
Changing the integration momenta $l\leftrightarrow k-l$, we obtain
\[
\widetilde{I}(n_1,n_2,n_4,n_3;n_0)
=(-1)^{n_0}\widetilde{I}(n_1,n_2,n_3,n_4;n_0) \; .
\]
In particular, when $n_3=n_4$ the integrals with odd $n_0$ vanish.
Integrals with $n_{3,4}\le0$ or $n_1\le0$ are trivial, as before.

In principle, one could deal with (\ref{tildeI}) by generalizing
(to higher $n_0$) the substitutions like 
$p\cdot l\to (p\cdot k)(k\cdot l)/k^2$ (cf.\ eq.~(\ref{r1})).
However, the higher $n_0$ is, the more cumbersome these
substitutions are. An alternative, recursive way is to use 
the integration-by-parts procedure\footnote{A similar algorithm has been
constructed in~\cite{AK}.}.
Applying $\frac{\partial}{\partial k}\cdot p$ to the integrand,
we obtain the recurrence relation
\begin{equation}
n_4\4+\0+ \widetilde{I} =
\Bigl[ -n_1+n_2 - M^2 n_0\0- + n_1(\2-+2M^2)\1+ - n_2\1-\2+
- \tfrac{1}{2}n_4(\1--\2-)\4+ \Bigr] \widetilde{I}\,.
\label{n1}
\end{equation}
We may assume $n_4\ge n_3$;
if $n_4>1$, then both $n_0$ and $n_4$ can be reduced by~(\ref{n1}).
Thus we are left with already known integrals
and $\widetilde{I}(n_1,n_2,1,1;n_0)$ with even $n_0$.
Applying $\frac{\partial}{\partial k}\cdot l$ to the integrand,
we obtain at $n_3=n_4=1$
\begin{eqnarray}
&&n_1\1+\0+ \widetilde{I}(n_1,n_2,1,1;n_0)
\nonumber\\
&{}={}&\Bigl[ - \tfrac{1}{4}n_0(\1--\2-)\0-
+ 2n_1\3-\1+ + 2n_2\3-\2+ + (\2--\3-+\!2m^2)\4+ \Bigr]
\widetilde{I}(n_1,n_2,1,1;n_0)\,.
\hspace{7mm}
\label{n2}
\end{eqnarray}
If $n_1>1$, then both $n_0$ and $n_1$ can be reduced by~(\ref{n2}).
Now, we are left with known integrals
and $\widetilde{I}(1,n_2,1,1;n_0)$ with even $n_0$.
Finally, applying $\frac{\partial}{\partial l}\cdot l$
to the integrand, we obtain at $n_1=n_3=n_4=1$
\begin{equation}
(d+n_0-3)\widetilde{I}(1,n_2,1,1;n_0) =
- (\2--\3-+4m^2)\4+ \widetilde{I}(1,n_2,1,1;n_0)\,.
\label{n3}
\end{equation}
The right-hand side contains integrals with the same $n_0$
but $n_4>1$; in them, $n_0$ can be again reduced by~(\ref{n1}).
In this way, these more general integrals~(\ref{tildeI})
are also reduced to the basis ones, $I_{0,1}$ and $T_{0,1}$.

\section{Effect of $m_c$ on $b$ quark chromomagnetic interaction}
\label{Sec:CMag}

First, we calculate contributions of a flavour with mass $m$
(e.g., the $c$ quark)
to the mass and wave function renormalization constants
of a quark with mass $M$ 
(e.g., the $b$ quark)
in the on-shell scheme (Fig.~\ref{Fig:Self})
in $d$ dimensions.
The results are\footnote{In QCD, $T_F=\frac{1}{2}$,
$C_F=(N^2-1)/(2N)$, $C_A=N$, where
$N=3$ is the number of colours.}
\begin{equation}
\label{Z_M}
\Delta Z_M = T_F C_F \frac{g_0^4}{(4\pi)^d} \Biggl[
- \frac{(d\!-\!2)^2(d\!-\!5)}{(d\!-\!3)(d\!-\!6)M^4}
\left(\frac{T_0}{2r^2}\!+\!T_1\right)
- \frac{d^2\!-\!9d\!+\!16}{(d\!-\!6)M^2} I_0
+ \frac{4(d\!-\!5)(1\!+\!r^2)}{d\!-\!6} I_1 \Biggr],
\end{equation}
\begin{equation}
\label{Z_Q}
\Delta Z_Q = T_F C_F \frac{g_0^4}{(4\pi)^d}
\frac{d-1}{1-r^2} \sum_{i=1}^4 a_i J_i\,,
\end{equation}
where
\begin{eqnarray*}
&&J_1 = \frac{(d-2)^2 T_0}{(d-3)(d-5)(d-6)M^2 m^2}\,,\quad
  J_2 = \frac{(d-2)^2 T_1}{(d-1)(d-3)(d-6)M^4}\,,\\
&&J_3 = \frac{I_0}{(d-1)(d-6)M^2}\,,\quad
J_4 = \frac{I_1}{(d-1)(d-6)}\,.
\end{eqnarray*}
The results for the coefficients $a_i$ in $\Delta Z_Q$ are
\begin{eqnarray*}
a_1 &{}={}& (2 d^{4} r^{4}-d^{4} r^{2}-39 d^{3} r^{4}+21 d^{3} r^{2}
-2 d^{3}+272 d^{2} r^{4}-159 d^{2} r^{2}+30 d^{2}
\\
&&{}-789 d r^{4}+509 d r^{2}-144 d+770 r^{4}-566 r^{2}+216)
/\bigl[4(d-2)(d-7)r^{2}\bigr]\,,
\\
a_2 &=& (d^{2}+5 d r^{2}-12 d-25 r^{2}+31)/2\,,
\\
a_3 &=& (2 d^{3} r^{2}+d^{3}-17 d^{2} r^{2}-12 d^{2}+27 d r^{2}
+47 d+8 r^{2}-56)/2\,,
\\
a_4 &=& 2 (d^{3} r^{2}-12 d^{2} r^{2}+d^{2}-5 d r^{4}+45 d r^{2}
-6 d+25 r^{4}-54 r^{2}+5) \,.
\end{eqnarray*}
After expansion in $\ep$ up to finite terms,
these formulae reproduce the results of~\cite{GBGS,BGS}.

In the case $m=0$,
\begin{equation}
T_0 = T_1 = 0\,,\quad
I_1 = - \frac{3d-8}{4(2d-7)} M^2 I_0\,.
\label{m0}
\end{equation}
Note that $I_0$ at $m=0$ is called $-M^{2d-6}I_1$ in~\cite{CG}.
$\Delta Z_Q$ contains a term $J_1/r^2\sim T_0/m^4$
which does not vanish in the limit $m\to0$.
This means that 
there is no smooth $m\to0$ limit in $\Delta Z_Q$~\cite{BGS}.
Omitting the $T_{0,1}$ terms and then setting $m=0$,
we reproduce the results of~\cite{GBGS,BGS}
for the massless quark contributions, exactly in $d$ dimensions.

In the case $m\to M$,
\begin{eqnarray}
T_1 &{}={}& T_0 \left[1+\frac{d-2}{2}(1-r^2)\right]
+ \mathcal{O}\bigl((1-r^2)^2\bigr)\,,
\nonumber\\
I_1 &=& - \frac{3d-8}{4(d-4)M^2} I_0 \left[1+\frac{1-r^2}{2}\right]
- \frac{3(d-2)^2}{8(d-3)(d-4)M^4} T_0
\left[1+\frac{2d+3}{6}(1-r^2)\right]
\nonumber\\
&&{} + \mathcal{O}\bigl((1-r^2)^2\bigr)\,.
\label{m1}
\end{eqnarray}
Note that $T_0$ and $I_0$ at $m=M$
are called in~\cite{CG} $M^{2d-4}I_0^2$ and $-M^{2d-6}I_2$,
respectively. 
Using this, we reproduce the results of~\cite{GBGS,BGS}
for the contributions of the quark with mass $M$, 
exactly in $d$ dimensions.

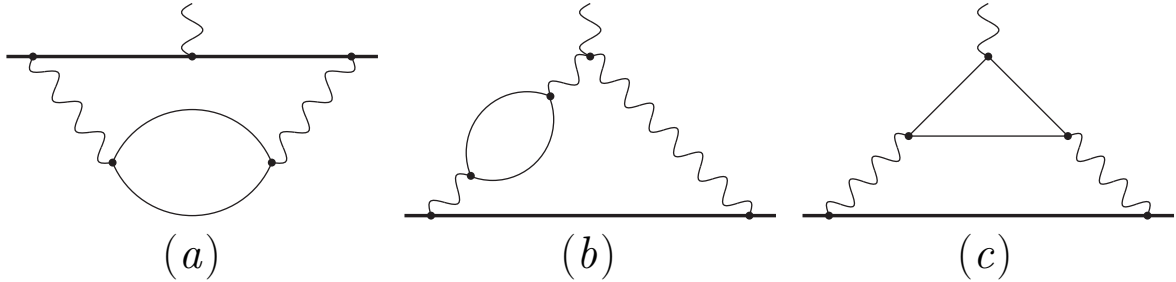
\begin{figure}[ht]
\begin{center}
\begin{picture}(440,90)
\SetWidth{1.3}
\Line(0,70)(140,70)
\SetWidth{0.5}
\Photon(70,70)(70,90){4}{1.5}
\Photon(10,70)(40,30){-4}{3.5}
\Photon(100,30)(130,70){-4}{3.5}
\CArc(70,42.5)(32.5,202.62,337.38)
\CArc(70,17.5)(32.5,22.62,157.38)
\Vertex(10,70){1.5}
\Vertex(130,70){1.5}
\Vertex(70,70){1.5}
\Vertex(40,30){1.5}
\Vertex(100,30){1.5}
\Text(70,-14)[b]{\Large(\textit{a})}
\SetWidth{1.3}
\Line(150,10)(290,10)
\SetWidth{0.5}
\Photon(220,70)(220,90){4}{1.5}
\Photon(220,70)(280,10){4}{6.5}
\Photon(160,10)(175,25){4}{1.5}
\Photon(205,55)(220,70){4}{1.5}
\CArc(196.25,33.75)(22.981,67.62,202.38)
\CArc(183.25,46.25)(22.981,247.62,22.38)
\Vertex(160,10){1.5}
\Vertex(280,10){1.5}
\Vertex(220,70){1.5}
\Vertex(175,25){1.5}
\Vertex(205,55){1.5}
\Text(220,-14)[b]{\Large(\textit{b})}
\SetWidth{1.3}
\Line(300,10)(440,10)
\SetWidth{0.5}
\Photon(370,70)(370,90){4}{1.5}
\Line(340,40)(370,70)
\Line(370,70)(400,40)
\Line(400,40)(340,40)
\Photon(310,10)(340,40){4}{3.5}
\Photon(400,40)(430,10){4}{3.5}
\Vertex(310,10){1.5}
\Vertex(430,10){1.5}
\Vertex(340,40){1.5}
\Vertex(400,40){1.5}
\Vertex(370,70){1.5}
\Text(370,-14)[b]{\Large(\textit{c})}
\end{picture}
\end{center}
\caption{Diagrams for the chromomagnetic moment:
(\textit{b}) implies also the mirror symmetric diagram,
and (\textit{c}) has two orientations of the quark loop.}
\label{Fig:CMag}
\end{figure}

Now we calculate the contribution $\Delta\mu_{\mathrm{g}}$
of a flavour with mass $m$ (say, $c$)
to the chromomagnetic moment of a quark with mass $M$ (say, $b$)
(Fig.~\ref{Fig:CMag}),
taking into account the wave function renormalization~(\ref{Z_Q}).
We use the background field method~\cite{Ab}.
Discontinuity in the limit $m\to0$ in the $C_F$ term
has cancelled with $Z_Q$, but in the $C_A$ term it is still here.
We reproduce the $m=0$ result (by omitting $T_{0,1}$ terms) and 
the $m=M$
result from Appendix in~\cite{CG}, exactly in $d$ 
dimensions\footnote{There is a typo in the Appendix of~\cite{CG}:
$J_{0,1,2}$ should read $J_{1,2,3}$}.

Combining this result with the corresponding HQET term
$\Delta\widetilde{\mu}_{{\mathrm{g}}0}$~\cite{CG}, we obtain
\begin{equation}
\Delta\mu_{\mathrm{g}}-\Delta\widetilde{\mu}_{{\mathrm{g}}0} =
T_F \frac{g_0^4}{(4\pi)^d} \frac{1}{1-r^2}
\left[ C_F \sum_{i=1}^4 a_{Fi}J_i
+ C_A \sum_{i=1}^4 a_{Ai}J_i \right]\,,
\label{mu}
\end{equation}
with the coefficients
\begin{eqnarray*}
a_{F1} &{}={}& (d^{3} r^{2}-11 d^{2} r^{2}-3 d^{2}+33 d r^{2}
+30 d-19 r^{2}-71)/2\,,
\\
a_{F2} &=& -(d^{3} r^{2}-2 d^{3}-17 d^{2} r^{2}+27 d^{2}+89 d r^{2}
-116 d-133 r^{2}+151)\,,
\\
a_{F3} &=& d^{4} r^{2}+2 d^{4}-7 d^{3} r^{2}-31 d^{3}-7 d^{2} r^{2}
+168 d^{2}+105 d r^{2}-375 d-152 r^{2}+296\,,
\\
a_{F4} &=& 4 (d^{4} r^{2}+d^{3} r^{4}-15 d^{3} r^{2}-17 d^{2} r^{4}
+85 d^{2} r^{2}-d^{2}+89 d r^{4}-221 d r^{2}+6 d
\\
&&{}-133 r^{4}+210 r^{2}-5)\,,
\\
a_{A1} &=& -(d^{3}-d^{2} r^{2}-13 d^{2}+11 d r^{2}+52 d-26 r^{2}-64)
/8\,,
\\
a_{A2} &=& (2 d^{3} r^{2}-3 d^{3}-29 d^{2} r^{2}+39 d^{2}
+129 d r^{2}-156 d-182 r^{2}+200)/4\,,
\\
a_{A3} &=& -(3 d-8)(d^{3}+d^{2} r^{2}-11 d^{2}-11 d r^{2}+38 d
+26 r^{2}-44)/4\,,
\\
a_{A4} &=& -(d^{4} r^{2}+2 d^{3} r^{4}-18 d^{3} r^{2}+2 d^{3}
-29 d^{2} r^{4}+114 d^{2} r^{2}-18 d^{2}
\\
&&{}+129 d r^{4}-299 d r^{2}+44 d-182 r^{4}+282 r^{2}-28)\,.
\end{eqnarray*}
Discontinuity in the limit $m\to0$ in the $C_A$ term has cancelled.
The difference of this contribution and that of a massless flavour
is finite as $\ep\to0$;
it represents the amount by which the $b$ quark chromomagnetic
interaction coefficient changes due to a non-zero mass of $c$ quark:%
\footnote{In the notations of~\cite{BG}, $A_F=\Delta_2-4\Delta_3$.}
\begin{eqnarray}
&&\Delta C_{\mathrm{m}} = T_F \left(\frac{\alpha_s}{4\pi}\right)^2
\left[ 8 C_F A_F + \frac{4}{3} C_A A_A \right]\,,
\label{Cm}\\
&&A_F = - r(1+r)(1-r-4r^2)L_+ + r(1-r)(1+r-4r^2)L_-
+ 6r^2 \left( \log r + \tfrac{4}{3} \right)\,,
\nonumber\\
&&A_A = - (1+r)(2+4r-r^2)L_+ - (1-r)(2-4r-r^2)L_-
+ 2\log^2 r + \tfrac{1}{3}\pi^2 + 2r^2 (\log r + 1)\,.
\nonumber
\end{eqnarray}
It vanishes as $r\to0$, whereas at $r=1$ it reproduces the result of
ref.~\cite{CG} (namely, the one with heavy quark loop).

In the QED case ($T_F=1$, $C_F=1$, $C_A=0$, $\alpha_s\to\alpha$),
the formula~(\ref{Cm}) is closely related to the contribution
of a lepton with mass $m$ (e.~g., electron or $\tau$)
to the magnetic moment of a lepton with mass $M$ (e.g., muon).
Adding back the massless contribution
(which was subtracted in~(\ref{Cm}))
and re-expressing the $\overline{\mathrm{MS}}$ coupling
in the one-loop contribution via the on-shell coupling
$\alpha(M)=\alpha\left(1-\frac{2\alpha}{3\pi}\log r\right)$,
we find that this contribution is given by~(\ref{Cm})
with $A_F\to A_F-\frac{2}{3}\log r-\frac{25}{18}$.
It was calculated in~\cite{El}; our result agrees with it
(and has a simpler form).

Expanding~(\ref{Cm}) at $r\ll1$ and $r\gg1$, we obtain
\begin{eqnarray*}
&&A_F = \tfrac{1}{2}\pi^2 r + 2 (4\log r + 3) r^2
- \tfrac{5}{2} \pi^2 r^3
+2\left(2\log^2r-\tfrac{14}{3}\log r+\tfrac{1}{3}\pi^2
+\tfrac{44}{9}\right)r^4
\\
&&\phantom{A_F={}} + \sum_{n=3}^{\infty} \left( 2g_F(2n) \log r
+ \frac{\mathrm{d}g_F(2n)}{\mathrm{d}n} \right) r^{2n}
= \sum_{n=0}^{\infty} \left( - 2g_F(-2n)\log r
+ \frac{\mathrm{d}g_F(-2n)}{\mathrm{d}n} \right) r^{-2n}\,,
\\
&&A_A = 3 \pi^2 r
+ 3 \left(\log^2 r + 4 \log r + \tfrac{1}{6} \pi^2 - 3 \right) r^2
- \tfrac{1}{2}\pi^2 r^3
+ \sum_{n=2}^{\infty} \left( 2g_A(2n) \log r
+ \frac{\mathrm{d}g_A(2n)}{\mathrm{d}n} \right) r^{2n}
\\
&&\phantom{A_A}
= 2\log^2 r + \tfrac{43}{3}\log r + \tfrac{1}{3}\pi^2
+ \tfrac{239}{18}
+ \sum_{n=1}^{\infty} \left( - 2g_A(-2n)\log r
+ \frac{\mathrm{d}g_A(-2n)}{\mathrm{d}n} \right) r^{-2n}\,,
\\
&&g_F(x) = \frac{1}{x-1} - \frac{5}{x-3} + \frac{4}{x-4}\,,\quad
g_A(x) = - \frac{2}{x} + \frac{6}{x-1} + \frac{3}{x-2}
- \frac{1}{x-3}\,.
\end{eqnarray*}
Note that loop effects of a very heavy flavour (say, $t$)
do not decouple in $C_{\mathrm{m}}$,
because it is not directly measurable.
Adding $\left(-\frac{2}{3}\log r-\frac{25}{18}\right)$
to the expansions of $A_F$, we obtain the expansions
of the contribution of a lepton with mass $m$
to the magnetic moment of a lepton with mass $M$ in QED;
they agree with the expansions of~\cite{El} obtained in~\cite{SL}.
It is also easy to obtain as many terms as needed in the expansions
as $r\to1$ using the formulae for $L_-$ via $1-r$ or $1-r^{-1}$
(see eq.~(\ref{L-}))
and for $L_++L_-$ via $1-r^2$ or $1-r^{-2}$ (see eq.~(\ref{LL})).

\section{Conclusions}
\label{Sec:Conc}

We have presented an algorithm which reduces
any scalar integral $\widetilde{I}(n_1,n_2,n_3,n_4;n_0)$
(given by eq.~(\ref{tildeI}))
to two trivial basis integrals~(\ref{T})
and two nontrivial ones, whose $\ep$-expansions
up to the finite terms are given by eqs.~(\ref{I0}) and (\ref{I1}).
The algorithm has been implemented as a \textsf{REDUCE} package.
It has been tested by evaluating about 27000 recurrence
relations for specific values of $n_i$, including those relations
which were not directly used for construction of the algorithm.
This package allows one to calculate, completely automatically, the 
on-shell two-loop self-energy
diagrams (Fig.~\ref{Fig:Self}) in any theory with two massive
particles having different masses\footnote{The package 
may be obtained from \textsf{http://www.inp.nsk.su/$\sim$grozin/}.}.

We calculated the effect of non-zero $m_c$ on the $b$ quark
chromomagnetic interaction coefficient~(\ref{Cm}).
Combining it with the result for $m_c=0$~\cite{CG},
we have numerically
\begin{equation}
C_{\mathrm{m}}(m_b) \simeq 1 + \frac{13}{6}\frac{\alpha_s(m_b)}{\pi}
+ (14.1439 + \Delta_{\mathrm{m}}(m_c/m_b))
\left(\frac{\alpha_s}{\pi}\right)^2\,,\quad
\Delta_{\mathrm{m}}(0.3\pm0.03) = 0.98\pm0.07\,.
\label{num}
\end{equation}
The effect is moderate and positive.
This is the only nontrivial coefficient in the HQET Lagrangian
up to $1/m_b$ level.

Due to reparametrization invariance~\cite{LM},
it also determines~\cite{CKO} the spin-orbit interaction coefficient
--- the most important spin-symmetry breaking coefficient
at $1/m_b^2$:
\begin{equation}
C_{\mathrm{so}}(\mu) = 2C_{\mathrm{m}}(\mu) -1\,.
\end{equation}
This relation has a very simple physical interpretation.%
\footnote{AGG thanks I.~B.~Khriplovich for this interpretation.}
The spin-orbit interaction is the interaction of a moving
(nonrelativistic) heavy quark with chromoelectric field.
In the quark rest frame,
the field acquires a chromomagnetic component.
The quark chromomagnetic moment interacts with it,
producing the term $2C_{\mathrm{m}}$.
There is also Thomas precession, which compensates half
of the chromomagnetic contribution at the tree level.
This is a purely kinematic effect due to Lorentz transformations,
and hence it gets no corrections.

The influence of $m_c$ on other $1/m_b^2$ terms in the $b$ quark
HQET Lagrangian, as well as on $1/m_b$ expansions
of various QCD currents,
can also be calculated using the presented method.

\textbf{Acknowledgements.}
We are grateful to K.~Melnikov and J.~B.~Tausk for communicating
unpublished details of~\cite{CM,FT} and for fruitful discussions,
to D.~J.~Broadhurst, A.~Czarnecki, M.~Yu.~Kalmykov and A.~V.~Kotikov
for comments on the manuscript,
and to J.~G.~K\"{o}rner for his hospitality in Mainz during 
the final stage of the work.
AID's research was essentially supported
by the Alexander von Humboldt Foundation, and partly
by the EU grant INTAS--93-0744.
AGG's work was supported in part by the BMBF
under contract No.\ 06~MZ865.

\section*{Appendix~A. Integrals in $d$ dimensions}

Here we present the most general results for the integrals~(\ref{I}).
To indicate that the results of this appendix are valid for
arbitrary powers of propagators, not only the integer ones,
we substitute $n_i\to \nu_i$. 

Using Mellin-Barnes contour integral representation for 
the massive denominator ${\mathcal{D}}_1^{-\nu_1}$ 
(for details, see in~\cite{MB}), we get
\begin{eqnarray*}
I(\nu_1,\nu_2,\nu_3,\nu_4) = 
\frac{M^{d-2\nu_1-2\nu_2}\; m^{d-2\nu_3-2\nu_4}}
     {\Gamma(\nu_1) \Gamma(\nu_3) \Gamma(\nu_4)} \;
\frac{1}{2\pi \mbox{i}} 
\int\limits_{-\mathrm{i}\infty}^{\mathrm{i}\infty}
\mathrm{d}s \left( \frac{M^2}{m^2} \right)^{\!s} \;
\Gamma(-s) \; \Gamma\left( \nu_1\!+ \nu_2\! - \tfrac{d}{2}-\!s\right)
\\
\times \frac{\Gamma(\nu_3+s) \; \Gamma(\nu_4+s) \;
             \Gamma\left(\nu_3+\nu_4-\tfrac{d}{2}+s \right) \;
             \Gamma(d-\nu_1-2\nu_2+2s)}
            {\Gamma(\nu_3+\nu_4+2s) \; \Gamma(d-\nu_1-\nu_2+s)}\,,
\end{eqnarray*}
where the integration contour separates right and left series
of poles of $\Gamma$ functions occurring in the numerator of
the integrand.

If we close the contour to the right, we should sum over two series 
of poles\footnote{If we close the contour to the left, we get 
a more cumbersome result
in terms of the hypergeometric series of $m^2/M^2$, corresponding
to the analytic continuation of (\ref{5F4}).}. 
The general result can be presented as
\begin{eqnarray}
I(\nu_1,\nu_2,\nu_3,\nu_4) = 
(m^2)^{d-\nu_1-\nu_2-\nu_3-\nu_4}
\hspace{95mm}
\nonumber\\[2mm]
\times \left\{
\frac{\Gamma\left( \frac{d}{2}-\nu_1\!-\nu_2 \right) \;
      \Gamma( \nu_1\!+\!\nu_2\!+\!\nu_3\!+\!\nu_4\!-d) \;
      \Gamma\left( \nu_1\!+\! \nu_2\!+\! \nu_3\!- \frac{d}{2} \right) \;
      \Gamma\left( \nu_1\!+\! \nu_2\!+\! \nu_4\!- \frac{d}{2} \right)}
     {\Gamma(\nu_3) \; \Gamma(\nu_4) \; \Gamma\left( \frac{d}{2} \right)\;
      \Gamma( 2\nu_1 + 2 \nu_2 + \nu_3 + \nu_4 -d)}
\right. \hspace{9mm}
\nonumber\\[1mm]
\times\left. _5F_4\left(\;
\begin{array}{c}
\nu_1\!+\!\nu_2\!+\!\nu_3\!-\!\tfrac{d}{2}, \;
\nu_1\!+\!\nu_2\!+\!\nu_4\!-\!\tfrac{d}{2}, \;
\nu_1\!+\!\nu_2\!+\!\nu_3\!+\!\nu_4\!-\!d, \;
\tfrac{1}{2}\nu_1, \;
\tfrac{1}{2}(\nu_1\!+\!1) \\[1mm]
\nu_1\!+\!\nu_2\!+\!\tfrac{1}{2}(\nu_3\!+\!\nu_4)
   \!-\!\tfrac{d}{2}, \;
\nu_1\!+\!\nu_2\!+\!\tfrac{1}{2}(\nu_3\!+\!\nu_4\!+\!1)
   \!-\!\tfrac{d}{2}, \;
\nu_1\!+\!\nu_2\!-\!\tfrac{d}{2}\!+\!1, \;
\tfrac{d}{2}
\end{array}\; \right|
\frac{M^2}{m^2} \right)
\nonumber\\[2mm]
+ \left(\frac{M^2}{m^2} \right)^{\tfrac{d}{2}-\nu_1-\nu_2} \;
\frac{\Gamma\left( \nu_1+\nu_2-\tfrac{d}{2}\right) \;
      \Gamma\left( \nu_3 +\nu_4-\tfrac{d}{2}\right) \;
      \Gamma(d-\nu_1-2\nu_2)}
     {\Gamma(\nu_1) \; \Gamma(\nu_3+\nu_4) \; \Gamma(d-\nu_1-\nu_2)}
\hspace{32mm}
\nonumber\\[1mm]
\left.
\times \left._5F_4\left(\;
\begin{array}{c}
\nu_3, \; \nu_4, \; \nu_3+\nu_4-\tfrac{d}{2}, \;
\tfrac{1}{2}(d-\nu_1-2\nu_2), \;
\tfrac{1}{2}(d-\nu_1-2\nu_2+1) \\[1mm]
\tfrac{1}{2}(\nu_3+\nu_4), \;
\tfrac{1}{2}(\nu_3+\nu_4+1), \;
d-\nu_1-\nu_2, \; 
\tfrac{d}{2}-\nu_1 -\nu_2+1 
\end{array}\; \right|
\frac{M^2}{m^2} \right)
\right\}\,,\hspace{5mm}
\label{5F4}
\end{eqnarray}
where
\[
\left.
_PF_Q\left(\;\begin{array}{c} a_1, \ldots , a_P \\
                            c_1, \ldots , c_Q
\end{array}\; \right| z \right)
= \sum_{j=0}^{\infty} \frac{(a_1)_j \ldots (a_P)_j}
                           {(c_1)_j \ldots (c_Q)_j}\;
\frac{z^j}{j!}  
\]
is the generalized hypergeometric function;
$(a)_j\equiv \Gamma(a+j)/\Gamma(a)$ is the Pochhammer symbol.

For trivial cases $\nu_1=0$ and $\nu_3=0$ (or $\nu_4=0$),
the expression (\ref{5F4})
reproduces known results in terms of $\Gamma$ functions.
When $\nu_2=0$ (the ``sunset'' configuration),
the same result~(\ref{5F4}) can be reproduced
in this limit (i.e., $p^2=M^2$ and two other masses are equal) 
by taking two sums in the hypergeometric series  
of three variables presented in ref.~\cite{BBBS}.   

One can see that for integer (non-negative) values of
$\nu_i$ both $_5F_4$ functions in (\ref{5F4}) reduce,
for arbitrary $d$, to a finite sum of $_3F_2$ functions of the
same argument. For example, we get for our basis integrals
$I_0=I(1,0,1,1)$ and $I_1=I(1,1,1,1)$:
\begin{eqnarray*}
\frac{I_0}{\Gamma^2(1+\ep)}=-\frac{(m^2)^{1-2\ep}}{\ep^2(1-\ep)}
\left\{ \frac{1}{1-2\ep}\; \left.
_3F_2\left(\;\begin{array}{c} 1, \tfrac{1}{2}, -1+2\ep \\
                            2-\ep, \tfrac{1}{2}+\ep
\end{array}\;
\right| \frac{M^2}{m^2} \right)
\right. 
\\
\left.
 + \left(\frac{M^2}{m^2}\right)^{1-\ep}\; \left.
_3F_2\left(\;\begin{array}{c} 1, \ep, \tfrac{3}{2}-\ep \\ 
                            3-2\ep, \tfrac{3}{2}
\end{array}\; \right| \frac{M^2}{m^2} \right)
\right\} \, ,
\end{eqnarray*}
\begin{eqnarray*}  
\frac{I_1}{\Gamma^2(1+\ep)}=-\frac{(m^2)^{-2\ep}}{\ep^2}
\left\{ \frac{1}{2(1-\ep)(1+2\ep)}\; \left.
_3F_2\left(\;\begin{array}{c} 1, \tfrac{1}{2}, 2\ep \\ 
                            2-\ep, \tfrac{3}{2}+\ep
\end{array}\; \right| \frac{M^2}{m^2} \right)
\right.
\\
\left.
 - \left(\frac{M^2}{m^2}\right)^{-\ep}\; \frac{1}{1-2\ep} \; \left.
_3F_2\left(\;\begin{array}{c} 1, \ep, \tfrac{1}{2}-\ep \\
                            2-2\ep, \tfrac{3}{2}    
\end{array}\; \right| \frac{M^2}{m^2} \right)
\right\} \, ,
\end{eqnarray*}
exactly in $d=4-2\ep$ dimensions\footnote{If $M=m$, one arrives 
at a representation in terms of $_3F_2$
functions of unit argument~\cite{B}.}.
Expansion of $I_0$ and $I_1$ in $\ep$ up to finite terms is given in 
eqs.~(\ref{I0})--(\ref{I1}).

\section*{Appendix~B. Various ways to calculate basis integrals}

Using a procedure similar to~\cite{FT}, the result~(\ref{I1})
can be derived as follows. 
Note that $m^{4\ep}I_1$ and $m^{4\ep}I(2,0,1,1)$ are dimensionless
and depend only on $r=m/M$.
Differentiating the definition~(\ref{I}) in $M$
and taking into account that $p=Mv$ ($v^2=1$),
we see that $I_1$ should obey the differential equation\footnote{This
equation can also be derived using the method of~\cite{Ko}.}
\begin{equation}
r^2 \frac{\mathrm{d}}{\mathrm{d}r}
\left[ r^{-1}\; m^{4\ep}\; I_1 \right]
= - m^{4\ep} I(2,0,1,1)\,.
\label{difeq}
\end{equation}
Therefore,
\begin{equation}
\label{intsol}
m^{4\ep}\; I_1
= r\;\int\limits_{r}^{\infty} \frac{\mathrm{d}\rho}{\rho^2}\;
m^{4\ep}\left.I(2,0,1,1)\right|_{r\to\rho}\,.
\end{equation}
The expression for $I(2,0,1,1)$ can be taken, e.g., from
ref.~\cite{BDU}:
\begin{equation}
\label{I2011} 
\frac{I(2,0,1,1)}{\Gamma^2(1+\ep)}
= (M^2)^{-2\ep} 
\left[
\frac{1}{2\ep^2}
+ \frac{1}{2\ep} - \frac{1}{2}
- 2 \log^2r   
+ 2(1-r^2) \left(L_+ + L_-\right)
\right] + \mathcal{O}(\ep)\,.
\end{equation}
Substituting eq.~(\ref{I2011}) into~(\ref{intsol}) we get~(\ref{I1}).

An alternative derivation follows the method used in~\cite{CM}.
The integrals (\ref{I}) can be presented as
\begin{equation}
I(n_1,n_2,n_3,n_4)=-\frac{\mathrm{i}}{\pi^{d/2}}\left.
\int \frac{\Pi_{12}(k^2)\,\mathrm{d}^d k}
          {\mathcal{D}_1^{n_1}\mathcal{D}_2^{n_2}}\right|_{p^2=M^2},
\;\;\;\; \mbox{with} \;\;\;\;
\Pi_{n_3 n_4}(k^2) \equiv -\frac{\mathrm{i}}{\pi^{d/2}} \int
\frac{\mathrm{d}^d l}{\mathcal{D}_3^{n_3} \mathcal{D}_4^{n_4}}\,.
\end{equation}
For instance, when we consider integrals with $n_3=1$ and $n_4=2$
we get 
\begin{equation}
\Pi_{12}(k^2) = \frac{\lambda^2-1}{4m^2 \lambda}
\log\frac{\lambda+1}{\lambda-1} + \mathcal{O}(\ep),
\;\;\;\; \mbox{with} \;\;\;\;
\lambda\equiv\sqrt{1-\frac{4m^2}{k^2}} \, .
\label{Pi12}
\end{equation}
Moreover, we know that $\Pi_{12}(k^2)$ may be represented via
dispersion integral,
$\Pi_{12}(k^2)=\int\mathrm{d}s \rho(s)/(s-k^2)$, 
the explicit form of the spectral density $\rho(s)$ being not
relevant for our discussion.

As the first example, let us consider the integral $I(2,0,1,2)$,
which is convergent. Substituting the spectral representation 
for $\Pi_{12}$, combining the denominators $\mathcal{D}_1^2$
and $(s-k^2)$ by the Feynman formula and, then, calculating
the integrals over $k$ and $s$, we arrive at
\begin{equation}
I(2,0,1,2) = \int_0^1
\mathrm{d}x \; \frac{x}{1-x}\;
\left. \Pi_{12}(k^2)\right|_{k^2=-\frac{M^2 x^2}{1-x}}
+\mathcal{O}(\ep)\,.
\label{I2012_} 
\end{equation}
If we consider, say,
$I(3,0,1,2)$, we get in the integrand
$\left.\left(\mathrm{d}\Pi_{12}(k^2)/\mathrm{d}k^2\right)
\right|_{k^2=-M^2x^2/(1-x)}$.
The second example is $I(2,1,1,2)$. In this case, the on-shell 
singularity can be separated via $\Pi_{12}(k^2)=
\left[\Pi_{12}(k^2)\!-\!\Pi_{12}(0)\right]+\Pi_{12}(0)$.
Note that 
$\left[\Pi_{12}(k^2)\!-\!\Pi_{12}(0)\right]/k^2
=\int\mathrm{d}s \rho(s)/[s(s\!-\!k^2)]$.
In this way, we obtain
\begin{equation}
\frac{I(2,1,1,2)}{\Gamma^2(1+\ep)} =
- \frac{(Mm)^{-2-2\ep}}{4\ep}
- \int_0^1 \mathrm{d}x \; \frac{x}{1-x}\,
\left.\frac{\Pi_{12}(k^2)-\Pi_{12}(0)}{k^2}
\right|_{k^2=-\frac{M^2 x^2}{1-x}}
+ \mathcal{O}(\ep)\,.
\label{I2112}
\end{equation}
Integrals with $n_2>1$ require more subtractions,
and those with $n_1>2$ involve derivatives.

After inserting the expression (\ref{Pi12}), the integrals over $x$
are of the following form:
\begin{equation}
\int_0^1 \left[ \frac{A(x)}{\lambda(x)} 
\log\frac{\lambda(x)+1}{\lambda(x)-1} + B(x) \right]
\mathrm{d}x\, ,
\label{fey}
\end{equation}
where $\lambda(x)\equiv\left.\lambda\right|_{k^2=-M^2x^2/(1-x)}$
(cf.\ eq.~(\ref{Pi12})), whilst
$A(x)$ and $B(x)$ are rational functions.
Introducing the new variable $y$ (the limits of the $y$ integration
are also from 0 to 1),
\begin{equation}
\label{y}
y=\frac{x(\lambda(x)-1)}{2r},
\;\;\;\; \mathrm{so}\;\;\mathrm{that} \;\;\;\;
x=\frac{r(1-y^2)}{y+r}
\;\;\; \mathrm{and} \;\;\;
\frac{1}{\lambda(x)}\frac{\mathrm{d}x}{\mathrm{d}y}=
-\frac{r(1-y^2)}{(y+r)^2}\,,
\end{equation}
we see that all occurring structures become rational, since
\begin{equation}
\lambda(x)=\frac{1+2ry+y^2}{1-y^2},
\hspace{10mm}
\frac{\lambda(x)+1}{\lambda(x)-1}=\frac{1+ry}{y(r+y)} \,.
\end{equation}
Therefore, the integral (\ref{fey}) can be expressed
in terms of dilogarithms.
In this way, from~(\ref{I2012_}), (\ref{I2112}) we obtain
\begin{equation}
I(2,0,1,2) = \frac{L_+\!+\!L_-}{M^2} + \mathcal{O}(\ep)\,,\quad
\frac{I(2,1,1,2)}{\Gamma^2(1+\ep)} = \frac{1}{4M^2 m^{2+4\ep}}
\left[ - \frac{1}{\ep} + r(L_+\!-\!L_-) + 2 \right]
+ \mathcal{O}(\ep)\,.
\label{I2012}
\end{equation}
This is enough to reproduce~(\ref{I0}), (\ref{I1}).

Similar procedure can be also applied to $I(n_1,n_2,n_3,n_4)$
with other values of $n_i$,
as soon as a suitable subtraction can be performed.
Apart from subtracting $\Pi_{n_3n_4}(0)$ (and higher terms of the
Taylor expansion in $k^2$), subtractions of 
$\left.\Pi_{n_3n_4}\right|_{m=0}$ may be also considered.

Some further results of interest are
\begin{eqnarray*}
&&I(1,0,2,2) = -2\frac{L_++L_-}{M^2}-\frac{L_+-L_-}{m^2}
+ \mathcal{O}(\ep)\, ,
\\
&&I(2,0,2,2) = \frac{1}{4M^2 m^2}\left[ r(L_+ - L_-)+2
-\frac{2r^2\log r}{1-r^2}\right]
+ \mathcal{O}(\ep)\, .
\end{eqnarray*}
In particular, if we take $I(2,0,1,2)$ and $I(1,0,2,2)$ as
independent finite integrals, we get only ``pure'' 
$L_+\pm L_-$ combinations, without logarithms, etc. 

The results~(\ref{I0}) and~(\ref{I1}) can also be checked by using
the formulae of Appendix~A in~\cite{BG}.
The quantities $\Delta_{1,2,3}$ were calculated there, which are
certain combinations of the integrals
\begin{eqnarray}
&&(M^2)^{2-n_1-n_2} J(n_1,n_2)
\equiv - \frac{\mathrm{i}}{\pi^{d/2}}
\int \frac{P(k^2,m)-P(k^2,0)}
{\mathcal{D}_1^{n_1}\mathcal{D}_2^{n_2}} \mathrm{d}^d k
= \left.I(n_1,n_2,1,1)\right|_{m=0}
\nonumber\\
&&\quad{} - I(n_1,n_2,1,1)
+ 2I(n_1,n_2+1,1,0) + \frac{4m^2}{d\!-\!2} I(n_1,n_2+1,1,1) \,  ,
\label{J}
\end{eqnarray}  
where
\begin{eqnarray}
\label{P}
P(k^2,m) &{}={}& -\left(1+\frac{4m^2}{(d-2)k^2}\right)\Pi_{11}(k^2)
-\frac{2}{k^2}\Pi_{10}
\nonumber \\
&=& m^{-2\ep} \Gamma(1+\ep) \left[ - \frac{1}{\ep}
+ \left(1+\frac{2m^2}{k^2}\right)
\left(\lambda\log\frac{\lambda+1}{\lambda-1}-2\right) \right]
+ \mathcal{O}(\ep) 
\end{eqnarray}
is proportional to the fermion-loop contribution to 
the gluon polarization operator\footnote{In refs.~\cite{GBGS,BG}
the notation 
$\Pi(-m^2/k^2)\equiv\left[P(k^2,m)-P(k^2,0)\right]_{d=4}$
has been used.
To get the $d$-dimensional unrenormalized contribution 
to the gluon polarization operator, eq.~(\ref{P}) should be
multiplied by $-2T_Fg_0^2(4\pi)^{-d/2}(d-2)/(d-1)$.}.
When an integral contains neither ultraviolet nor infrared (on-shell)
singularities, it can be calculated in four dimensions.
Averaging $1/\mathcal{D}_1^{n_1}$ over the directions of $k$
in four-dimensional Euclidean space,
we obtain $A_{n_1}/\mathcal{D}_2^{n_1}$,
where $A_0=A_{-1}=1$, $A_1=x$, $A_2=x^2/(2-x)$,
and $k^2=-M^2 x^2/(1-x)$.
Therefore, convergent integrals are given by
\begin{equation}
\label{Jn1n2}
J(n_1,n_2) = \int_0^1 \Phi_{n_1 n_2}(x)
\left[P(k^2,m)-P(k^2,0)\right]_{k^2=-\frac{M^2 x^2}{1-x}}
\mathrm{d}x\,,
\end{equation}
\[
\mbox{with} \hspace{7mm}
\Phi_{n_1 n_2}(x) = A_{n_1}(x) x^{3-2n_1-2n_2} (1-x)^{n_1+n_2-3}
(2-x)\,.
\]
In terms of these integrals, the quantities $\Delta_{1,2,3}$
used in~\cite{BG} can be expressed as
\begin{equation}
\label{Delta}
\Delta_1=\tfrac{1}{6}\left[2J(1,1)\!+\!J(2,0)\right], \;\;
\Delta_2=\tfrac{1}{3}\left[J(1,1)\!-\!J(2,0)\right], \;\;
\Delta_3=\tfrac{1}{6}\left[J(0,1)\!-\!J(1,0)\!-\!J(2,0)\right]\,.
\end{equation}
This coincides with the integral representations~\cite{BG}.
Explicit expressions for $\Delta_i$ given in eq.~(A2) of~\cite{BG}
(which can be obtained using the substitution~(\ref{y})
provide three independent checks on our basis integrals.
Note that the integral calculated in~\cite{GBGS}
is $\Delta=\frac{1}{4}(\Delta_1+\Delta_3)$.
  
The integrals\footnote{An integral of this type was considered in 
\cite{BGS}, where the notation 
$\bar{\Pi}(k^2)\equiv\left[P(k^2,m)-P(0,m)\right]_{d=4}$
was used.}
\begin{eqnarray}
&&(M^2)^{2-n_1-n_2} \bar{J}(n_1,n_2)
\equiv - \frac{\mathrm{i}}{\pi^{d/2}}
\int \frac{P(k^2,m)-P(0,m)}
{\mathcal{D}_1^{n_1}\mathcal{D}_2^{n_2}} \mathrm{d}^d k
= \frac{2(d-1)}{3(d-2)} I(n_1,n_2,2,0)
\nonumber\\
&&\quad{} - I(n_1,n_2,1,1)
+ 2I(n_1,n_2+1,1,0) + \frac{4m^2}{d-2} I(n_1,n_2+1,1,1)
\label{Jbar}
\end{eqnarray}
have better infrared convergence but worse ultraviolet one.
We define $\bar{\Delta}_{1,2,3}$ via
\begin{eqnarray}
&&\bar{J}(2,1) = \bar{\Delta}_1\,,\quad
\bar{J}(2,0)-\bar{J}(1,1) = \bar{\Delta}_2\,,
\\
&&\bar{J}(0,1)-\bar{J}(1,0)-\bar{J}(1,1) =
\bar{\Delta}_3 + \frac{\ep}{2-\ep} \bar{J}(0,2)\,.
\label{Deltabar}
\end{eqnarray}  
The combination of integrals on the left-hand side
of (\ref{Deltabar}) is rather interesting.
As $k\to\infty$, the combination of their denominators behaves as
\[
\frac{1}{M^2\mathcal{D}_2} - \frac{1}{M^2\mathcal{D}_1}
- \frac{1}{\mathcal{D}_1\mathcal{D}_2} \Rightarrow
\frac{\ep}{(2-\ep)\; k^4} + \mathcal{O}\left(\frac{1}{k^6}\right)\,,
\]
where (on the right-hand side) terms vanishing when averaged 
over the directions of $k$ are omitted.
The term $\mathcal{O}(1/k^6)$ gives
an ultraviolet-convergent integral.
The term $\mathcal{O}(\ep/k^4)$ might have been omitted,
if an arbitrarily large ultraviolet cutoff were considered.
Without the cutoff, it gives a purely ultraviolet contribution
(similar to the axial anomaly) $\ep\bar{J}(0,2)/(2-\ep)$.
The term $\bar{\Delta}_3$ can be calculated at $d=4$
using the convergent integral representation with
$\Phi_{01}(x)-\Phi_{10}(x)-\Phi_{11}(x)$.
Using eq.~(\ref{y}), we obtained
\begin{eqnarray*}
&&\bar{\Delta}_1 = \tfrac{1}{4}(1+r)(4-r+r^2)L_+
+ \tfrac{1}{4}(1-r)(4+r+r^2)L_-
+ \left(\tfrac{5}{3}+\tfrac{1}{2}r^2\right)\log r
+ \tfrac{14}{9} + \tfrac{1}{2}r^2\,,
\\
&&\bar{\Delta}_2 = 3r(1-r^2)(L_+-L_-) + 2(2-3r^2)\log r
+ \tfrac{16}{3} - 6r^2\,,
\\
&&\bar{\Delta}_3 = 3r(1+r)^2(1-2r)L_+ - 3r(1-r)^2(1+2r)L_-
- 3(1-4r^2)\log r + \tfrac{19}{4} - 15r^2
\end{eqnarray*}
and checked that these expressions agree with $d$-dimensional results
obtained by our program as $d\to4$.
The integral calculated in~\cite{BGS} is $\bar{\Delta}=
\frac{1}{16}(4\bar{\Delta}_1+3\bar{\Delta}_2-2\bar{\Delta}_3)$.
Eqs.~(\ref{I0}) and~(\ref{I1}) can be reconstructed from any two
out of four independent integrals from~\cite{GBGS,BGS,BG}.

\end{document}